\documentclass[12pt,a4paper]{article}

\usepackage{graphicx}
\usepackage{dcolumn}
\usepackage{bm}
\usepackage{hyperref}
\usepackage{color}
\usepackage{latexsym,bm}
\usepackage{amsmath,amsfonts}
\usepackage{amssymb}
\usepackage{mathtools}
\usepackage{mathrsfs}
\usepackage{float}
\usepackage{graphicx}
\usepackage[lofdepth,lotdepth]{subfig}
\usepackage[square,numbers, sort&compress]{natbib}

\usepackage{dcolumn}
\usepackage[mathlines]{lineno}
\usepackage{geometry}
\usepackage[affil-it]{authblk}  
\usepackage{color}
\usepackage[toc,page]{appendix}
\renewcommand{\vec}[1]{\mathbf{#1}}
\allowdisplaybreaks[4] 
\begin{document}

\title{
Path-averaged Kinetic Equation for Stochastic Systems  \footnote{This study is supported by NSFC with Grant No. 91547204}
}

\author[1,$\dagger$]{De-yu ZHONG}
\author[2]{Guang-qian WANG}
\author[3]{Tie-jian LI}
\author[4]{Ming-xi ZHANG}
\author[5]{You XIA}
\author[6]{Yu ZHANG}
\affil[1,2,3,4,5,6]{State Key Laboratory of Hydroscience and Engineering 
\\Tsinghua University, Beijing 10084, China }
\affil[$\dagger$]{Corresponding author: zhongdy@tsinghu.edu.cn }




\date{\today}


\maketitle

\begin{abstract}
For a stochastic system, its evolution from one state to another can have a large number of possible paths. Non-uniformity in the field of system variables leads the local dynamics in state transition varies considerably from path to path and thus the distribution of the paths affects statistical characteristics of the system. Such a characteristic can be referred to as path-dependence of a system, and long-time correlation is an intrinsic feature of path-dependence systems. 
We employed a local path density operator to describe the distribution of state transition paths, and based on which we derived a new kinetic equation for path-dependent systems.  
The kinetic equation is similar in form to the Kramers-Moyal expansion, but with its expansion coefficients determined by the cumulants with respect to state transition paths, instead of transition moments. This characteristic makes it capable of accounting for the non-local feature of systems, which is essential in studies of large scale systems where the path-dependence is prominent.  Short-time correlation approximation is also discussed. It shows that the cumulants of state transition paths are equivalent to jump moments when correlation time scales are infinitesimal, as makes the kinetic equation derived in this paper has the same physical consideration of the Fokker-Planck equation for Markov processes. 
\end{abstract}

\maketitle

%
%
\section{\label{sec:level1}Introduction}


Complex systems, for instance, flows from small scale diffusion \cite{heppe_1998} to large scale geophysical currents \citep{Berloff2005Random}, are often found to exhibit a high degree of uncertainty, which makes deterministic prediction a difficult task. In addition to sophisticated work on improving deterministic prediction from every angle, statistical mechanical approaches are developed \citep{balescu,risken1984fokker,van1992stochastic, zwanzig2001nonequilibrium} and becomes one of the prominent theories to cope with uncertainties as an intrinsic feature of complex systems \citep{Berloff2005Random, Non-Statitic-Geo-Fluid, Zidikheri2010Stochastic,Dentz2010Probability,MINIER20011,Kang2013Uncertainty}.   

Usually, evolution of a stochastic system is decomposed into two parts, a deterministic one and stochastic one; as a common way, the stochastic part is assumed Markovian, so that it is described readily by the well-know Fokker-Planck equation (FPE) \citep{risken1984fokker}. Statistical characteristics of a stochastic process described by FPE is determined by its state transition moments, the averages of powers of state differences between two successive states over all possible transitions \citep{risken1984fokker}. Because the state transition moments rely completely on the state differences in the phase space of two succesive states over a small time interval, the Markovian FPE is unable to consider long-time correlation \citep{risken1984fokker}, an essential feature found in systems with multiple time scales. There are a number of  improvements made on the Markovian FPE to account for long-time correlation problems \citep{Adelman1976Fokker,Faetti1988,van1992stochastic,zwanzig2001nonequilibrium,Sokolov2002Solutions,Santamar2003Memory,Mura2008Non,Ilyin2012Fokker,Sandev2015Diffusion, MALTBA201887}.


The memory effect arising from long-time correlation can be viewed from a different angle. The states of a system are described by its general coordinates as a set of points on the phase space and thus each of the variation in the state marks a path leading from one point to another in the phase space. On the one hand, if the variation in a system is of strong rate-dependence on internal or external system variables, memory effect becomes significant and presents itself along the entire state transition paths on the phase space, possessing a typical non-Markovian nature. On the other hand, random forcing, from internal fluctuations of a system as well as from that of its environment, makes its evolution from one state to another can have a large number of possible paths. Moreover, non-homogeneity in the field of system variables can make the local dynamic behaviors in state transition differs significantly from path to path. Apparently, the distribution of state transition paths affect ultimately the statistical properties of the system. Such a relationship between statistical properties of a system and the probability distribution of state transition paths can be referred to as path-dependence, of which memory effect is intrinsic.

The path-dependence is a general property of systems with multiple time scales and long-time correlation in state transition, in which the time rates of changes of the system variables in states are so significant that they cannot be neglected any more, and cannot be estimated linearly as the differences between adjacent states either. For instance, in studies of turbulent flows by means of statistical mechanics, macro flows are regarded as collections of motion of eddies with different temporal and spatial scales. It has been demonstrated that motion of large eddies is usually non-local \citep{SBPOPE-TURBULENTFLOWS,Tsuji2012}, resulting in that all actions that eddies experience along the different paths of state transition affect ultimately the statistical properties of turbulent flows. Such a property is not peculiar to turbulent flows, but occurs in many systems with multiple time scales.

The Markovian FPE focuses on a small segment of a path for state transition during a short time interval, and thus fails to account for the influences that take place all along the entire path due to long-time correlation.  In other words, from a point of view of statistical mechanics, the ensemble average defined in the Markovian FPE is on the states of a system and their statistical distribution; differently, the ensemble for a path-dependent stochastic process is required to be defined as a set of the state variables along with their evolution paths and corresponding distributions.  Apparently,  classical formulation of Markovian FPE applies well if a stochastic process  depends only on the changes of state variables; on the contrary, when  a system relies strongly not only on the changes of its state variables, but also on its transition paths, or more specifically, if the problems of interest are path-dependent, then Markovian FPE is insufficient. 
Therefore, how to explicitly take into account multiple possibilities in the paths of each transition step of a non-equilibrium system becomes crucial; especially, when we focus our attention to the effect on the evolution of a system arising from uncertainties in transition paths. This paper is aimed to provided a  formulation of  a  kinetic equation for path-dependent systems.

From the point of view of statistical mechanics, state transition paths together with their distribution constitute a statistical ensemble, by which we can formulate a new kinetic equation for path-dependent systems. Certainly, this requires a-priori knowledge about the paths, but it brings about no more essential difficult than FPE, except that we need to develop a new mechanism to identify the paths.  In fact, in many practical problems, state transition paths can be observed which helps us have a general view of statistical properties about  the state transition paths. In this study, we employed a local path density operator to determine the distribution of state transition paths. In the following sections, we firstly give a detailed derivation of the kinetic equation, then its approximation for  short time correlation problems is discussed, and followed by a simple example of the present study. Concluding remarks are presented in the last. This study shows that the kinetic equation derived in this paper is similar in form to the Kramers-Moyal expansion, but with the expansion coefficients expressed in terms of the cumulants with respect state transition paths, rather than jump moments. This characteristic makes it capable of accounting for the non-local feature of systems, which is essential in studies of large scale systems where path-dependence is prominent.

%
%
\section{\label{sec:level2}Formulation}
%
%
\subsection{\label{sec:level2-1} Local path density operator}
In statistical mechanics, a macroscopic state is regarded to correspond to a huge number of different microscopic configurations compatible with their macroscopic constraints \cite{balescu}, which leads any possible realization of the macroscopic state, as well as its transition paths in the phase space, to exhibits a certain degree of uncertainty. This uncertainty, especially that observed in state transition paths,  is what we concerned in this study. In order to formulate this uncertainty and its impact on the distribution of macroscopic states of a system, we must define a distribution function for state transition paths to constitute a statistical ensemble.

Considering a system described by the state variable $\vec{X}=\{{X}_j\}$  with $N$ components, and each component $X_j$ ($j=1,\cdots, N$) can be either a vector or a scalar depending on specific problems we are interested in. The path for state transition is elaborated as a curve on the phase space along which system state changes from $\vec{X}(s)$ at the time $s$ to arrive at $\vec{X}(t)$ at the time $t$ and denoted as $\vec{X}(t)=\vec{X}(t|\vec{X}(s),s)$. Furthermore, we assume that $\vec{X}(t)$ is differentiable, or at least piecewise differentiable, with respect to time $t$, and it observes that 
\begin{eqnarray}\label{eq-1}
\dot{\vec{X}}\equiv\frac{\mathrm{d}\vec{X}}{\mathrm{d} t}   = H(\vec{X}), 
\end{eqnarray}
where $H$ is an arbitrary integrable function of the variable $\vec{X}$. In addition,  we assume the first order system of Eq. \eqref{eq-1} has an initial state (value) of  
\begin{eqnarray}\label{eq-2}
\vec{X}(s)   = \vec{X}(s|\vec{X}(s),s).
\end{eqnarray}

Considering a special case that the system is at the state $\vec{x}$ at the time $t$, conditioned it is in the state of $\vec{y}=\{{y}_{1}, {y}_{2},\cdots,{y}_{N}\}$ at the time $s$, i.e., $\vec{X}(s)=\vec{y}$. In this case, the state transition curve is $\vec{X}(t)=\vec{X}(t|\vec{y},s)$ with $\vec{X}(s|\vec{y},s)=\vec{y}$.  
In order to identify those paths arriving at $\vec{x}=\{x_1,\cdots x_N\}$ at the time $t$, we introduce a local (fine-grained) operator $\chi$ which describes the density of state transition paths that passes $\vec{y}$ at the time $s$ to $\vec{x}$ at the time $t$ given by Eq. \eqref{eq-1}. It is:
\begin{eqnarray}\label{eq-3}
\chi(t)  &=& \chi(|\vec{x}-\vec{X}(t|\vec{y},s)|) 
=\prod_{j=1}^N \chi(|x_j-X_j(t|\vec{y},s)|). 
\end{eqnarray}
%
%
The local path density operator $\chi$ is assumed to have a sharp value at $\vec{x}=\vec{X}(t|\vec{y},s)$ while it is zero elsewhere. The most simple choice of $\chi$ is the Dirac-$\delta$ function, but for the sake of generality, we still keep it as a general functional of the state difference $|\vec{x}-\vec{X}(t|\vec{y},s)|$ as it in Eq. {\eqref{eq-3}}.

From the definition of the local path density operator, we inferred that $ \chi$ depends on the path $\vec{X}(t)=\vec{X}(t|\vec{y},s)$ which change with time. The time rate of change of $\chi $ along the curve $\vec{X}(t|\vec{y},s)$ is
\begin{eqnarray}\label{eq-4}
\frac{\partial \chi(|\vec{x}-\vec{X}(t|\vec{y},s)|)}{\partial {t}} 
=
\dot{\vec{X}} \nabla_{\vec{X}} \chi(|\vec{x}-\vec{X}(t|\vec{y},s)|).
\end{eqnarray}
Since $\chi=\chi(|\vec{x}-\vec{X}(t|\vec{y},s)|)$ is a function of $|\vec{x}-\vec{X}(t|\vec{y},s)|$, it is straightforward to verify the identity $\nabla_{\vec{X}} \chi=-\nabla_{\vec{x}}\chi$. Denoting $\mathscr{L}=\dot{\vec{X}}\nabla_{\vec{x}}=\sum_{j=1}^N{\dot{X}_j\partial/\partial x_j}$, Eq. \eqref{eq-4} is written in a compact form as  
\begin{eqnarray}\label{eq-5}
\frac{\partial \chi}{\partial {t}}=-\mathscr{L} \chi.
\end{eqnarray}

Eq. \eqref{eq-5} is a Liouville equation for local path density operator $\chi$, which has an operator solution along the path $\vec{X}(t)=\vec{X}(t|\vec{y},s)$ as \citep{van1992stochastic, zwanzig2001nonequilibrium, path-integrals}
\begin{eqnarray}\label{eq-6}
\chi(|\vec{x}-\vec{X}(t|\vec{y},s)|)  
&=&\mathscr{U}(t|s) \chi(|\vec{x}-\vec{X}(s|\vec{y},s)|) \nonumber \\
&=&\mathscr{U}(t|s)\chi(|\vec{x}-\vec{y}|), 
\end{eqnarray}
where $\mathscr{U}(t|s)$ is a time evolution operator defined by (see Refs.  \cite{van1992stochastic, path-integrals})
\begin{eqnarray}\label{eq-7}
\mathscr{U}(t|s)
&=&\overleftarrow{T}\mathrm{e}^{-\int_s^{{t}} \text{d} \tau \mathscr{L}(\tau)} \nonumber \\
&=&\overleftarrow{T}\sum_{n=0}^{\infty}\frac{(-1)^n}{n!} \left(\int_{s}^{t}\text{d}\tau
 \mathscr{L}(\tau) \right)^n,
\end{eqnarray} 
with $\overleftarrow{T}$ denoting  the time-ordering operator by which the integrations in
\begin{eqnarray}\label{eq-8}
\left(\int_{s}^{t} \mathrm{d}\tau \mathscr{L} \right)^n  
=
\int_{s}^{t} \mathrm{d}\tau_1  \cdots\int_{s}^{t}\text{d}\tau_{n}
\mathscr{L}(\tau_1)  \cdots \mathscr{L}(\tau_{n}), 
\end{eqnarray}
for $n=1, 2, \cdots$ are correctly ordered so that the earlier times in the products of the integrand stand to the left of those with later times ($\tau_{1}>\tau_{2}>\cdots>\tau_{n}$).

\subsection{\label{sec:level2-2} Probability density function}
In the previous studies, for a Markov process, the possibility to find the system having the macroscopic state of $\vec{x}$ in a coarse-grained level at the time $t$ is given by the Chapman-Kolmogorov equation, which maps the state $\vec{y}$ at the time $s$ to the state $\vec{x}$ at the time $t$ by means of the transition probability function $f(\vec{x},t|\vec{y},s)$. In the previously reported studies, the transition probability $f(\vec{x},t|\vec{y},s)$ is formally expanded as a Taylor series, which purely depends on changes of states and thus is a function of transition moments \citep{risken1984fokker}; meanwhile, it is limited to the cases that the jump time scale $|t-s|$ must be infinitesimal to make the transition moments exit. However, if we care about path-dependent stochastic processes, a statistical ensemble constitutes of paths through which systems change their microscopic states together with their distributions is necessary to derive a transition probability for path-dependent  stochastic processes.

A macroscopic state, in the context of statistical mechanics, can be regarded as function of micro configurations compatible with macroscopic constraints \citep{balescu}. Considering a  system is in the state $\vec{X}$ corresponding to a realization of the microscopic configuration $\Gamma$, denoted as  $\vec{X}(\Gamma)=\vec{X}(t|\vec{y},s; \Gamma)$.  Let $F(\Gamma)$ to be the distribution of the micro-state $\Gamma$, which satisfies $\int \mathrm{d} \Gamma  F(\Gamma) =1$, then $F(\Gamma) \mathrm{d}\Gamma$ is the possibility to find a system in the micro state of $(\Gamma, \Gamma+\mathrm{d}\Gamma)$. The ensemble average on the local path density function leads to the coarser-grained probability density to find the system at the state $\vec{x}$ at the time $t$, given that it is at the state $\vec{y}$ at the time $s$:
\begin{eqnarray}\label{eq-9}
f(\vec{x},t| \vec{y},s) 
&=&\int \mathrm{d} {\Gamma }F(\Gamma)
 \chi(|\vec{x}-\vec{X}(\Gamma)|) \nonumber\\
 &=&\langle \chi(|\vec{x}-\vec{X}(t|\vec{y},s)|)\rangle 
, 
\end{eqnarray}
where a variable closed by a pair of angles ``$\langle\rangle$" is its ensemble average defined by: 
\begin{eqnarray}\label{eq-10}
\langle A \rangle  
&=&
\int \text{d}\Gamma F(\Gamma) A(\Gamma).
\end{eqnarray} 

Eq. \eqref{eq-9} indicates that, the local path density operator $\chi$ assumes the function of picking up those paths leading from a given point $\vec{y}$ at the time $s$ to arrive at $\vec{x}$ at the time $t$ from all of state transition paths. This definition is different from classical statistical mechanics, in which a local density function is usually employed to pick up points of interest on the phase space \citep{balescu}. Nevertheless, both the local path density operator defined in this study and the local density function adopted in classical statistical mechanics are designed to obtain a coarser-grained distribution function.

By substituting Eq. \eqref{eq-6} into Eq. \eqref{eq-9}, the conditional probability density function $f(\vec{x},t| \vec{y},s)$ can be written in an equivalent form as: 
\begin{eqnarray} \label{eq-11}
f(\vec{x},t| \vec{y},s)
&=& 
\int \text{d} \Gamma  F(\Gamma)\chi(|\vec{x}-\vec{X}(t|\vec{y},s)|) \nonumber \\
&=& \langle \mathscr{U}(t|s) \rangle \chi(|\vec{x}-\vec{y}|),  
\end{eqnarray}
where $\langle \mathscr{U}(t|s) \rangle$ is the averaged time evolution operator and is expanded in detail as follows:
\begin{eqnarray}\label{eq-12}
\langle\mathscr{U}(t|s) \rangle 
&=&
\int \text{d}\Gamma  F(\Gamma) \mathscr{U}(t|s) \nonumber \\
&=& 
\mathrm{exp}
\left(
\overleftarrow{T}\sum_{n=1}^{\infty}\frac{(-1)^n}{n!}
\left\langle\left\langle\left(
\int_s^{t}\text{d}\tau \mathscr{L} \right)^n\right\rangle\right\rangle\right),
\end{eqnarray}
in which $\langle\langle\rangle\rangle$ represents the cumulant, for instance, $\langle\langle A \rangle\rangle= \langle A  \rangle$ and  $\langle\langle AB \rangle\rangle= \langle(A-\langle A  \rangle )(B-\langle B  \rangle ) \rangle$ are, respectively, first and second order cumulant regarding the ensemble average defined by Eq. \eqref{eq-5}. In the derivation of Eq. \eqref{eq-12}, we have used the result of the ensemble average of an exponent function \citep{risken1984fokker,van1992stochastic}.  Noticing that $f(\vec{x},s|\vec{y},s)=\chi(|\vec{x}-\vec{y}|)$, Eq. \eqref{eq-11} is a cumulant expansion of $f(\vec{x},t| \vec{y},s)$ at the time $s$; whereas the Kramers-Moyal expansion in deriving FPE is a moment expansion.  

$f(\vec{x},t| \vec{y},s)$ is a conditional distribution function for path-dependent systems. It is advisable to derive a probability density function for $\vec{x}$ in many circumstances. As usual, it can be obtained by the relation: 
\begin{eqnarray} \label{eq-13}
f(\vec{x},t)=\int \mathrm{d}\vec{y}f(\vec{x},t| \vec{y},s)f(\vec{y},s).  
\end{eqnarray}  
However, Eq. \eqref{eq-13} requires integration on the coarser-grained scale ($\vec{y}$ is a coarser-grained state), which makes the present study loss the capability to account for path-dependence on fine-grained scale as we expected. For this reason, it is necessary to define the distribution function for $\vec{x}$ through a different way.

Multiplying both sides of Eq. \eqref{eq-9} with a Dirac $\delta$-function of $\delta(|\vec{y}-\vec{X}(s)|)$, where $\vec{X}(s)$ is a known point on an arbitrary transition path $\vec{X}(t)=\vec{X}(t|\vec{X}(s),s)$  at the time $s$, we find that 
\begin{eqnarray} \label{eq-14}
\int{\mathrm{d}\vec{y}}f(\vec{x},t|\vec{y},s)\delta(|\vec{y}-\vec{X}(s)|) 
=\langle\chi(|\vec{x}- {\vec{X}}(t) |)\rangle.
\end{eqnarray}

\begin{figure}[htbp] 
\centering
\includegraphics[scale=0.45, trim=0cm 0.cm 0cm 0cm]{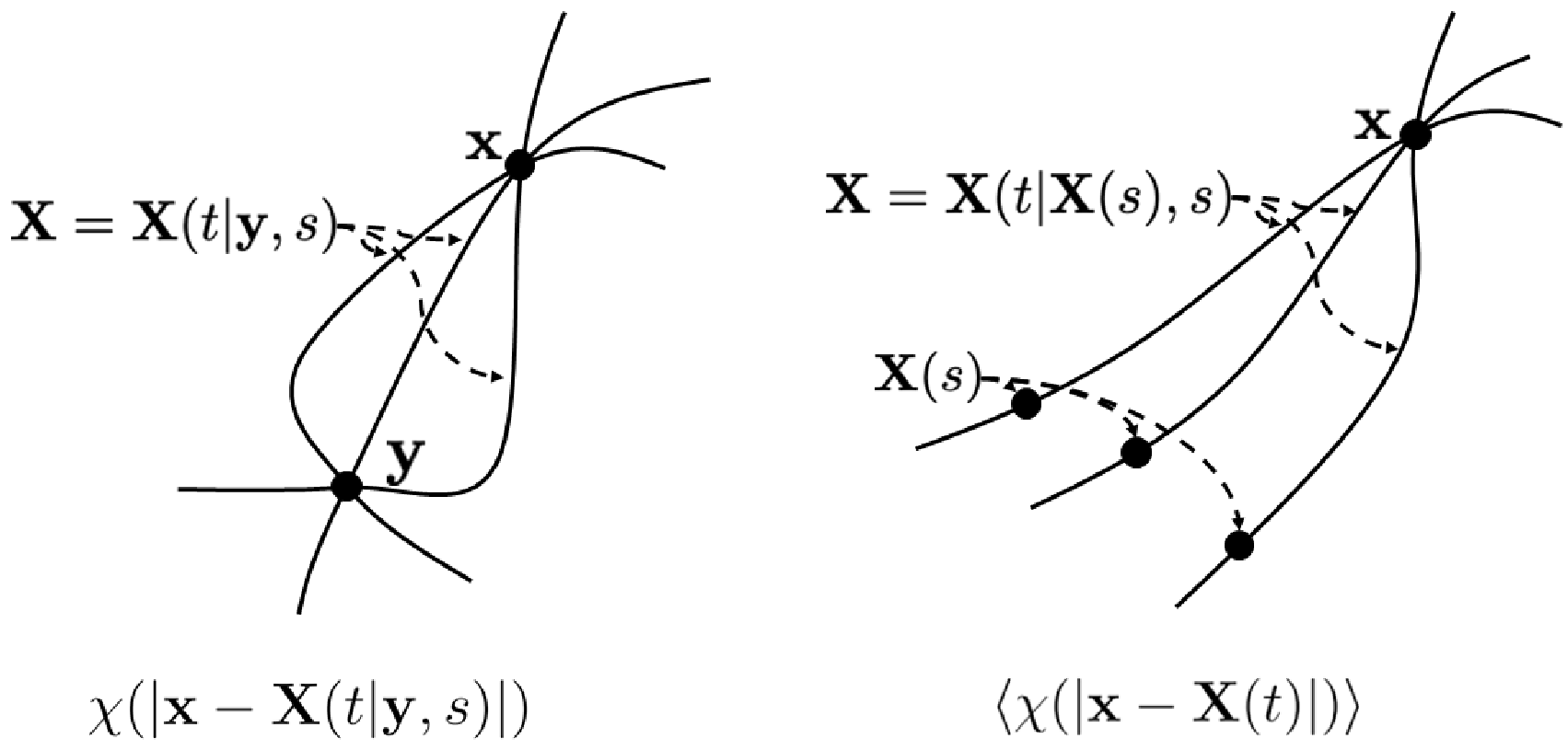} 
\caption{\label{fig:Fig1}{\emph{A schematic diagram of differences between $\langle \chi(|\vec{x}-\vec{X}(t|\vec{y},s)|)\rangle$ and $\langle\chi(|\vec{x}- {\vec{X}}(t) |)\rangle$.  It shows that $\langle \chi(|\vec{x}-\vec{X}(t|\vec{y},s)|)\rangle$ is an ensemble paths from a given state $\vec{y}$ at the time $s$ to arrive at $\vec{x}$ at the time $s$, while $\langle\chi(|\vec{x}- {\vec{X}}(t) |)\rangle$ is an ensemble taken on all of the paths leading to $\vec{x}$. Correspondingly, the time evolution operator $\langle \mathscr{U}(t|s)\rangle$ in Eq. \eqref{eq-11} involves the ensemble average on integral cures starting from a given state $\vec{y}$ of coarser-grained scale, while in Eq. \eqref{eq-15}, the ensemble average is taken on integral cures having an initial state $\vec{X}(s)$ of fine-grained scale.}}}
\end{figure}

Eq. \eqref{eq-14} helps us extend the focus on the state transition path from a specific point $\vec{y}$ to all the possible paths arriving at $\vec{x}$. The major differences between $\langle \chi(|\vec{x}-\vec{X}(t|\vec{y},s)|)\rangle$ and $\langle\chi(|\vec{x}- {\vec{X}}(t) |)\rangle$ are depicted in Fig. \ref{fig:Fig1}. It shows that, $\langle \chi(|\vec{x}-\vec{X}(t|\vec{y},s)|)\rangle$ is an ensemble average on the paths from a given state $\vec{y}$ at the time $s$ to arrive at $\vec{x}$ at the time $s$; whereas $\langle\chi(|\vec{x}- {\vec{X}}(t) |)\rangle$ is an ensemble average taken on all of the paths leading to $\vec{x}$. $ \langle\chi(|\vec{x}- {\vec{X}}(t) |)\rangle$ is in essence the distribution function for $\vec{x}$ at the time $t$, i.e., 
\begin{eqnarray} \label{eq-15}
f(\vec{x},t) &\equiv& \langle \chi(|\vec{x}- {\vec{X}}(t)|)\rangle 
=\langle \mathscr{U}(t|s) \rangle \chi(|\vec{x}-\vec{X}(s)|).  
\end{eqnarray}  
It must be noticed that, the time evolution operator $\langle \mathscr{U}(t|s)\rangle$ in Eq. \eqref{eq-11} involves the ensemble average on integral cures starting from a given state $\vec{y}$ of coarser-grained scale, while in Eq. \eqref{eq-15}, the ensemble average is taken on integral cures having an initial state $\vec{X}(s)$ of fine-grained scale.

%
%

\subsection{\label{sec:level2-3} Kinetic equation for path-dependent systems}
Differentiating Eq. \eqref{eq-11} with respect to  $t$,  inserting $\mathscr{L}=\dot{\vec{X}}\nabla_{\vec{x}}=\sum_{j=1}^N{\dot{X}_j\partial/\partial x_j}$ into the resulting equation, we derive a path-averaged kinetic equation for $f(\vec{x},t|\vec{y},s)$ including an infinite number of terms as follows:
\begin{eqnarray} \label{eq-16}
\frac{\partial f(\vec{x},t|\vec{y},s)}{\partial {t}} 
&=& 
\sum_{n=1}^{\infty}(-1)^n\nabla_{\vec{x}}^n   \langle\mathscr{D}^{(n)}(\vec{x},t| \vec{y},s) \rangle f(\vec{x},t|\vec{y},s),
\end{eqnarray}
where the coefficient $ \langle\mathscr{D}^{(n)}(\vec{x},t|\vec{y},s)\rangle $ is given by
\begin{eqnarray} \label{eq-17}
\langle \mathscr{D}^{(n)}(\vec{x},t|\vec{y},s)\rangle 
&=&
\frac{1}{n!}\frac{\partial}{\partial t}
\overleftarrow{T}
\left\langle\left\langle\left(
\int_s^{t}\text{d}\tau \dot{\vec{X}} \right)^n
\right\rangle\right\rangle.
\end{eqnarray}
The operator $\nabla_{\vec{x}}^n \langle \mathscr{D}^{(n)}(\vec{x},t|\vec{y},s)\rangle$ in Eq. \eqref{eq-16} is defined as    
\begin{eqnarray} \label{eq-18}
\nabla_{\vec{x}}^n \langle \mathscr{D}^{(n)}(\vec{x},t|\vec{y},s)\rangle 
&=&
\frac{\partial^n}{\partial x_{j_1}\cdots\partial x_{j_n}} \langle \mathscr{D}^{(n)}_{j_1\cdots j_n}(\vec{x},t|\vec{y},s)\rangle,
\end{eqnarray}
in which the summation convention with respect to repeated subscript $j_\nu$ ($j_{\nu}=1, 2, \cdots,  N$ and $\nu=1, \cdots, n$) is used, and the partial differential operators apply on all of the following variables.  

Furthermore, using the result given by Eq. \eqref{eq-14}, multiplying both sides of Eq. \eqref{eq-16} with $\delta(|\vec{y}-\vec{X}(s)|)$, integrating the resulting equation with respect to $\vec{y}$, we had a generalized kinetic equation for $f(\vec{x},t)$ as:  
\begin{eqnarray} \label{eq-19}
\frac{\partial f(\vec{x},t) }{\partial t} 
= 
\sum_{n=1}^{\infty}(-1)^n\nabla_{\vec{x}}^n  \langle \mathscr{D}^{(n)}(\vec{x},t) \rangle f(\vec{x},t),  
\end{eqnarray}
where the coefficient $\langle \mathscr{D}^{(n)}(\vec{x},t)\rangle$ involves the ensemble average on integral cures starting from an initial state $\vec{X}(s)$ on fine-grained scale, instead of a given state $\vec{y}$ on coarser-grained scale as in Eq. \eqref{eq-16}. Eq. \eqref{eq-19} can also be obtained by directly differentiating Eq. \eqref{eq-15}.

It shows that  Eq. \eqref{eq-16} (or Eq. \eqref{eq-19}) contains an infinite number of terms as the Kramers-Moyal expansion. According to the Pawula theorem, if the  Kramers-Moyal expansion does not truncated at the second order, it must contain an infinite number of terms \citep{risken1984fokker, Sto-m-cgispinGardiner}. This is true for the present study, although the expansion coefficients are function of cumulants \citep{Sto-m-cgispinGardiner}. 
When  Eq. \eqref{eq-19} is truncated at $n=2$, we had a kinetic equations as follows:
\begin{eqnarray} \label{eq-20}
\frac{\partial f(\vec{x},t) }{\partial t} 
+
\nabla_{\vec{x}} \langle \mathscr{D}^{(1)}(\vec{x},t) \rangle f(\vec{x},t) 
=
\nabla_{\vec{x}}^2 \langle \mathscr{D}^{(2)}(\vec{x},t) \rangle f(\vec{x},t),  
\end{eqnarray}
which resembles FPE derived by truncating the Kramers-Moyal expansion at the second order. However, it shows that the first order expansion coefficient $\langle\mathscr{D}^{(1)}\rangle$ is no longer a simple drift velocity; instead, it is a state transition velocity. Similarly, the second order expansion coefficient $\langle\mathscr{D}^{(2)}\rangle$ in not simply a diffusion coefficient but a measurement of dispersion of state transition paths. 
Because the significance of information contained by high-order cumulants deceases much quickly, unlike high-order moments which contain information about lower moments \citep{risken1984fokker, Sto-m-cgispinGardiner}, it is expected that the current study converges faster than the Kramers-Moyal expansion. Nevertheless, a generalized kinetic equation truncated at a higher order (for instance, $n\ge 3$) is useful in some special cases \citep{risken1984fokker, GNL-PLY}.

\subsection{\label{sec:level2-4} Expansion coefficient for the first order systems}
The coefficient $\langle \mathscr{D}^{(n)}\rangle$ in Eq. \eqref{eq-19} is a function of the integration of $\dot{\vec{X}}$ along the curve $C$: $\vec{X}(t)=\vec{X}(t|\vec{X}(s),s)$ leading from $\vec{X}(s)$ to $\vec{x}$. Denoting $\vec{S}$ to be the integral curve of the state transition path, then it is 
\begin{eqnarray}\label{eq-21}
\vec{S}  
=
\int_s^t \mathrm{d} \tau \dot{\vec{X}} = \int_C \mathrm{d} \vec{X}(t) 
=
\left\{\int_C \mathrm{d} {X}_1,\cdots \int_C \mathrm{d} {X}_N  \right\}. 
\end{eqnarray}
Eq. \eqref{eq-17} shows that $\langle\mathscr{D}^{(n)}\rangle$ is completely determined by the statistical information about the state transition paths $\vec{S}$. Therefore, if the distribution of the state transition paths are available, it can be solved in the same ways as that for FPE. However, the expression for $\langle\mathscr{D}^{(n)}\rangle$ is too formal to be used; we need to express it as an explicit function of state transition paths.

For the first order system given by Eq. \eqref{eq-1}, assuming that the $n^{th}$ order derivative of $H(\vec{y})$ exists  ($n \ge 1$),  then $H(\vec{X}(t))$ can be expanded at $\vec{y}$ along the path $\vec{X}(t)=\vec{X}(t|\vec{y},s)$ as 
\begin{eqnarray}\label{eq-22}
H(\vec{X}(t))&=&\sum_{n=0}^{\infty} \frac{\vec{S}^n(t)}{n!} \nabla^n_{\vec{y}} H(\vec{y})=\mathrm{e}^{\vec{S}(\tau)\nabla_{\vec{y}}}H(\vec{y}) 
=
\mathscr{Y}^{-}(\vec{y},\tau)H(\vec{y}),
\end{eqnarray}
where $\vec{S}(\tau)=\int_s^{\tau} \mathrm{d}\xi \dot{\vec{X}}(\xi)=\int_{C}\mathrm{d}\vec{X}$, and $\mathscr{Y}^{-}(\vec{y},\tau)= \mathrm{e}^{\vec{S}(\tau)\nabla_{\vec{y}}}$ is a backwards-shifting operator which pushes the state of the system from $\vec{X}(\tau)$ back to its initial state $\vec{y}$. Thus,  
\begin{eqnarray}\label{eq-23}
\langle\mathscr{D}^{(n)}(\vec{x},t|\vec{y},s)\rangle 
&=&
\frac{1}{n!}\frac{\partial}{\partial t}
\overleftarrow{T}
\left\langle\left\langle\left(
\int_s^{t}\text{d}\tau \dot{\vec{X}}(\tau) \right)^n
\right\rangle\right\rangle
\nonumber\\
&=&
\frac{1}{n!}\frac{\partial}{\partial t}
\overleftarrow{T}
\left\langle\left\langle\left(
\int_s^{t}\text{d}\tau H(\vec{X}(\tau))\right)^n
\right\rangle\right\rangle
\nonumber\\
&=&\frac{1}{n!}\frac{\partial}{\partial t}
\overleftarrow{T}
\left\langle\left\langle\left(
\int_s^{t}\text{d}\tau  \mathscr{Y}^{-}(\vec{y},\tau) H(\vec{y})\right)^n 
\right\rangle\right\rangle
.
\end{eqnarray}
As to $\langle \mathscr{D}^{(n)} (\vec{x},t)\rangle$, it is 
\begin{eqnarray}\label{eq-24}
\langle \mathscr{D}^{(n)} (\vec{x},t)\rangle 
=
\frac{1}{n!}\frac{\partial}{\partial t}
\overleftarrow{T}
\left\langle\left\langle\left(
\int_s^{t}\text{d}\tau  \mathscr{Y}^{+}(\vec{x},\tau) H(\vec{x})\right)^n 
\right\rangle\right\rangle,
\end{eqnarray}
where $\mathscr{Y}^{+}(\vec{x},\tau)=\mathrm{e}^{-\vec{S}(\tau)\nabla_{\vec{x}}}$ is a forwards-shifting operator with $\vec{S}(\tau)=\int_{\tau}^t \mathrm{d} \xi \dot{\vec{X}}(\xi)$. The differences between $\langle \mathscr{D}^{(n)} (\vec{x},t|\vec{y},s)\rangle$ and $\langle \mathscr{D}^{(n)} (\vec{x},t)\rangle$ is obvious and details can be found in section \ref{sec:level2-3}. 

Eq. \eqref{eq-23} (or \eqref{eq-24}) shows that $\langle \mathscr{D}^{(n)}\rangle$ is a function of the cumulants with respect to the integral curves of transition paths. For this reason, \eqref{eq-16} (or Eq. \eqref{eq-19}) can be referred to as the path-averaged kinetic equations. In addition, the ensemble average of integral cures $\langle \vec{S}\rangle =\int \mathrm{d} \Gamma F(\Gamma)\vec{S}(\Gamma)$ is essentially implemented on an ensemble composed of all possible state transition paths, and $\left\langle\left\langle {\vec{S}^n } \right\rangle\right\rangle $, the $n$th-order cumulants with respect $ \vec{S}  $, was also termed as the correlation function \citep{Sto-m-cgispinGardiner}; it can be interpreted, therefore, as a variable to reflect the statistical structure constructed by state transition paths of a system. Consequently, the expansion coefficient $\langle \mathscr{D}^{(n)}\rangle$ is the result of changes of the structure of state transition paths in the phase space. Expressing $\langle \mathscr{D}^{(n)}\rangle$ in terms of the correlation function $\left\langle\left\langle {\vec{S}^n } \right\rangle\right\rangle $ provides us with a new angle to view how a system evolves:  it is the variation in statistical structures of state transition paths that drives statistical properties of a system to evolve.

%
%
\section{\label{sec:level3}Short-time correlation approximation}

On the one hand, in this study, there is no limitation imposed on $t-s$, i.e., it is unnecessary to assume the time scales of the state transition to be infinitesimal, and therefore, the present study can be applied to those systems with long-time correlations. This is because the expansion coefficient as a function of the integral cure $\vec{S}=\int_\tau^t \mathrm{d} \xi \dot{\vec{X}}(\xi)$, which naturally contains the information along the state transition paths.

On the other hand, if our concern of the time interval $\Delta t =t-s$ is small, so that
\begin{eqnarray}\label{eq-25}
\lim_{\Delta t \to 0} \mathscr{Y}^{+}(\vec{x},\tau=t-\Delta t)
=\mathscr{Y}^{+}(\vec{x},t)=1,
\end{eqnarray}
by which we had that 
\begin{eqnarray}\label{eq-26}
\langle \mathscr{D}^{(n)} (\vec{x},t)\rangle 
&=&\lim_{\Delta t \to 0} \frac{1}{n!}\frac{\partial}{\partial t}
\overleftarrow{T}
\left\langle\left\langle\left(
\int_{t-\Delta t}^{t}\text{d}\tau  \mathscr{Y}^{+}(\vec{x},\tau) H(\vec{x})\right)^n 
\right\rangle\right\rangle
\nonumber \\
&=&\frac{1}{n!}\lim_{\Delta t \to 0}\frac{1}{\Delta t}
\left\langle\left\langle\left(\Delta t H(\vec{x}) \right)^n 
\right\rangle\right\rangle \nonumber \\
&=&\frac{1}{n!}\lim_{\Delta t \to 0}\frac{1}{\Delta t}
\left\langle\left\langle \Delta^n \vec{S}  
\right\rangle\right\rangle 
.
\end{eqnarray}
Therefore, for instance, for $n=1$ and $n=2$, we had, respectively, 
\begin{eqnarray}\label{eq-27}
\langle \mathscr{D}^{(1)}(\vec{x},t) \rangle
=\lim_{\Delta t \to 0} \frac{\langle \Delta \vec{S}\rangle}{\Delta t},
\end{eqnarray}
and assuming a long-term stable state is reached so that $\langle \dot{\vec{X}} \rangle \to 0 $,
\begin{eqnarray}\label{eq-28}
\langle \mathscr{D}^{(2)}(\vec{x},t) \rangle 
=\frac{1}{2} \lim_{\Delta t \to 0}\frac{ \langle \Delta^2 \vec{S} \rangle}{\Delta t}.
\end{eqnarray}

As a comparison, the coefficient of the Kramers-Moyal expansion, a function of  the transition moments, is given by \citep{risken1984fokker}:
\begin{eqnarray}\label{eq-29}
 \mathscr{D}^{(n)}  
=\frac{1}{n!}\lim_{\Delta t \to 0} \frac{\langle \Delta^n \vec{X} \rangle}{\Delta t} 
=\frac{1}{n!}\lim_{\Delta t \to 0} \frac{1}{\Delta t} \langle (\vec{X}(t+\Delta t)-\vec{X}(t)|_{=\vec{x}})^n \rangle,
\end{eqnarray} 
which gives that, for $n=1$  
\begin{eqnarray}\label{eq-30}
\mathscr{D}^{(1)} 
=\lim_{\Delta t \to 0} \frac{ \langle \Delta \vec{X} \rangle}{\Delta t} 
=\lim_{\Delta t \to 0} \frac{1}{\Delta t} \langle(\vec{X}(t+\Delta t)-\vec{X}(t)|_{=x}) \rangle, 
\end{eqnarray}
and for $n=2$
\begin{eqnarray}\label{eq-31}
\mathscr{D}^{(2)} 
=\lim_{\Delta t \to 0} \frac{ \langle \Delta^2 \vec{X} \rangle}{\Delta t}
=\frac{1}{2}\lim_{\Delta t \to 0} \frac{1}{\Delta t} \langle (\vec{X}(t+\Delta t)-\vec{X}(t)|_{=\vec{x}})^2 \rangle
,
\end{eqnarray}
with the ensemble average defined as $\langle A \rangle=\int \mathrm{d} \vec{X}(t+\Delta t)  A f(\vec{X}(t+\Delta t), t+\Delta t|\vec{x},t)$.

It shows that, when $\Delta t= t-s$ is infinitesimal to make $\Delta \vec{S} \approx \Delta \vec{X}$, and thus $\langle \mathscr{D}^{(1)}(\vec{x},t) \rangle$ and $\langle \mathscr{D}^{(2)}(\vec{x},t) \rangle$ are, respectively, similar in form to $\mathscr{D}^{(1)}$ and $\mathscr{D}^{(2)}$ for FPE, as indicates that the cumulation expansion with respect to state transition paths is reduced to moment expansion about state jumps for short-time correlation problems. This is easy to understand that, for an infinitesimal time interval $\Delta t$, the integral cure $\Delta \vec{S}$ can be well represented by a small segment $\Delta\vec{X}$, a linear approximation of the integral cures. 

Nevertheless, differences still exist. Firstly, the ensemble average in the present study is defined on the fine-grained level, while that for FPE is defined on the coarser-grained level. Secondly, the first order expansion coefficient $\langle \mathscr{D}^{(1)} \rangle$ (see Eq. \eqref{eq-27}) measures the state transition velocity averaged on all the possible state transition path passing $\vec{x}$ at the time $t$; similarly,  the second order expansion coefficient $\langle \mathscr{D}^{(2)} \rangle$ (see Eq. \eqref{eq-28}) is a measurement of dispersion degree of transition paths; large diffusion implies wide distribution of the transition paths. In comparison, the drift coefficient (Eq. \eqref{eq-30}) and diffusion coefficient (Eq. \eqref{eq-31}) in FPE are determined by the first and second order moments calculated on the coarse-grained level.

%
%
\section{\label{subsec:level4}A simple example}
Considering a simple one-dimensional linear case of Eq. \eqref{eq-1}:
\begin{eqnarray}\label{eq-32}
\frac{\mathrm{d} {X}(t)}{\mathrm{d}t}=-\mu ({X}(t)-{X}_e(t)), 
\end{eqnarray}
where $\mu$ is a positive constant; ${X}_e(t)$ is a system variable, which is regarded the equilibrium state of the system. Although Eq. \eqref{eq-32} is simple, yet it is a useful model equation representing important natural processes in alluvial systems \cite{fluvial-knighton,WUBAOSHENG-GEO}.

Eq. \eqref{eq-32} can be transformed into an equivalent form to simplify the following derivation. By substitution of ${Z}={X}-{X}_e$, Eq. \eqref{eq-32} is written as:  
\begin{eqnarray}\label{eq-33}
\frac{\mathrm{d} {Z}(t)}{\mathrm{d} t} = -\mu{Z}(t)- \dot{{X}}_e,
\end{eqnarray}
which help us to understand how the system given by \eqref{eq-32} to response to external excitations. 

Denoting $H(Z(t))=-\mu ({Z}(t)+\mu^{-1}\dot{{X}}_e)$ and expanding it at ${z}+\mu^{-1}\langle \dot{{X}}_e\rangle $, we derived the expansion coefficient $\langle \mathscr{D}^{(n)}\rangle$ for $n=1$ and $n=2$ as 
\begin{eqnarray}\label{eq-34}
\langle \mathscr{D}^{(1)}({z},t)\rangle 
&=&
\frac{\partial}{\partial t}
\left\langle
\int_s^{t}\text{d}\tau  \mathscr{Y}^{+}({z},\tau)  H({z}) 
\right\rangle \nonumber\\
&=&\langle \mathscr{Y}^{+}({z},t)  H({z}) \rangle =1 \cdot  H({z})
=-\mu {z}-\langle \dot{{X}}_e\rangle,
\end{eqnarray}
and
\begin{eqnarray}\label{eq-35}
\langle \mathscr{D}^{(2)}({z},t) \rangle 
&=&
\frac{1}{2!}\frac{\partial}{\partial t}
\overleftarrow{T}
\left\langle\left\langle\left(
\int_s^{t}\text{d}\tau \dot{{Z}}(\tau)\right)^2 
\right\rangle\right\rangle\nonumber \\
&=&
\frac{1}{2!}\frac{\partial}{\partial t}
\overleftarrow{T}
\left\langle\left\langle\left(
\int_s^{t}\text{d}\tau H({Z}(\tau))\right)^2 
\right\rangle\right\rangle\nonumber \\
&=&
\frac{1}{2!}\frac{\partial}{\partial t}
\overleftarrow{T}
\left\langle\left\langle\left(
\int_s^{t}\mathrm{d}\tau  \mathscr{Y}^{+}({z},\tau) H({z})\right)^2 
\right\rangle\right\rangle\nonumber \\
&\approx&\frac{(\nabla_{{z}}H({z}))^2}{2} \frac{\partial }{\partial  t} 
\overleftarrow{T}
\left\langle \left\langle\int_s^t \int_s^{t} \mathrm{d}\tau_1\mathrm{d}\tau_2 {S}(\tau_1)  {S}(\tau_2)\right\rangle \right\rangle, 
\end{eqnarray}
 respectively. In the derivation of Eq. \eqref{eq-35}, the higher order correlation terms are neglected. The autocorrelation of the state transition paths is found to be 
\begin{eqnarray}\label{eq-36}
\langle \langle{S}(\tau_1)  {S}(\tau_2)\rangle\rangle
&=&\langle \langle {Z}(\tau_1)  {Z}(\tau_2)\rangle \rangle
+
\mu^{-2}\langle \langle \dot{X}_e(\tau_1)  \dot{X}_e(\tau_2)\rangle \rangle\nonumber \\
&+&\mu^{-1}\langle \langle {Z}(\tau_1)  \dot{X}_e(\tau_2)\rangle \rangle
+\mu^{-1}\langle \langle \dot{X}_e(\tau_1)  {Z}(\tau_2)\rangle \rangle.
\end{eqnarray}
For the cases of  $\mu \tau_1\gg 1$ and $\mu \tau_2 \gg 1$, $\langle \langle{S}(\tau_1)  {S}(\tau_2)\rangle\rangle$ is approximated by the first term on the right-hand-side of Eq. \eqref{eq-36} as  
\begin{eqnarray}\label{eq-37}
\langle \langle{S}(\tau_1)  {S}(\tau_2)\rangle\rangle
&\approx &\langle \langle {Z}(\tau_1)  {Z}(\tau_2)\rangle \rangle\nonumber \\
&=&\int_s^{\tau_1} \int_s^{\tau_2} \mathrm{d}\tau_1^{\prime}\mathrm{d}\tau_2^{\prime}\mathrm{e}^{-\mu(\tau_1+\tau_2-\tau_1^{\prime}-\tau_2^{\prime})}\langle \langle\dot{{X}}_e(\tau_1^{\prime}) \dot{{X}}_e(\tau_2^{\prime})\rangle\rangle. 
\end{eqnarray}
In the last line of Eq. \eqref{eq-37}, the solution of Eq. \eqref{eq-33} for $Z(t)$ is used.  
If it is assumed that the time correlation of $\dot{{X}}_e-\langle \dot{{X}}_e \rangle$ is exponentially decayed, i.e., a colored noise given by 
\begin{eqnarray}\label{eq-38}
\langle \langle\dot{{X}}_e(\tau_1^{\prime}) \dot{{X}}_e(\tau_2^{\prime})\rangle\rangle=\nu {D} \mathrm{e}^{-\nu|\tau_1^{\prime}-\tau_2^{\prime}|},
\end{eqnarray}
where ${D}$ is the intensity of correlation and $\nu$ is the decay constant, then the time correlation function of the state transition paths is 
\begin{eqnarray}\label{eq-39}
\langle \langle{S}(\tau_1)  {S}(\tau_2)\rangle\rangle 
&\approx &\langle \langle {Z}(\tau_1)  {Z}(\tau_2)\rangle \rangle\nonumber \\
&=&\int_s^{\tau_1} \int_s^{\tau_2} \mathrm{d}\tau_1^{\prime}\mathrm{d}\tau_2^{\prime}\mathrm{e}^{-\mu(\tau_1+\tau_2-\tau_1^{\prime}-\tau_2^{\prime})}\langle \langle\dot{{X}}_e(\tau_1^{\prime}) \dot{{X}}_e(\tau_2^{\prime})\rangle\rangle\nonumber \\
&=&\nu D \mathrm{e}^{-\mu(\tau_1+\tau_2)} \int_s^{\tau_1} \int_s^{\tau_2} \mathrm{d}\tau_1^{\prime}\mathrm{d}\tau_2^{\prime}\mathrm{e}^{\mu(\tau_1^{\prime}+\tau_2^{\prime})} \mathrm{e}^{-\nu|\tau_1^{\prime}-\tau_2^{\prime}|}\nonumber \\
&=&2\nu D\mathrm{e}^{-\mu(\tau_1+\tau_2)} \int_s^{\tau_1} \int_s^{\tau_1^{\prime}} \mathrm{d}\tau_1^{\prime}\mathrm{d}\tau_2^{\prime}
\mathrm{e}^{(\mu-\nu)\tau_1^{\prime}+(\mu+\nu)\tau_2^{\prime}} \nonumber \\
&=&\frac{2\nu D\mathrm{e}^{-\mu(\tau_1+\tau_2)}}{\mu+\nu}
\int_s^{\mathrm{min}(\tau_1,\tau_2)}  \mathrm{d}\tau_1^{\prime}
\left[
\mathrm{e}^{(\mu+\nu)\tau_1^{\prime}}
-\mathrm{e}^{(\mu+\nu)s} 
\right]
\mathrm{e}^{(\mu-\nu)\tau_1^{\prime}} 
\nonumber \\
&\approx&\frac{\nu D}{\mu(\mu+\nu)}\mathrm{e}^{-\mu|\tau_1-\tau_2|}. 
\end{eqnarray}
Using Eq. \eqref{eq-39}, one arrives that  
\begin{eqnarray}\label{eq-40}
\langle \mathscr{D}^{(2)}({z},t) \rangle 
\approx
D\frac{\nu  }{\mu+\nu}(1-\mathrm{e}^{-\mu(t-s)}). 
\end{eqnarray}
Thus, the kinetic equation derived in this paper for Eq. \eqref{eq-33}, with the first two terms kept, is written as   
\begin{eqnarray}\label{eq-41}
\frac{\partial f({z},t)}{\partial t} 
=
\frac{\partial  (\mu z+\langle \dot{{X}}_e \rangle ) f({z},t)}{\partial {z}} 
+D\frac{\nu   \left(1-\mathrm{e}^{-\mu(t-s)}\right) }{\mu+\nu} \frac{\partial^2f({z},t)}{\partial {z}^2}. 
\end{eqnarray}
Eq. \eqref{eq-41} has a stationary solution for $\langle \dot{X}_e\rangle=0$ as 
\begin{eqnarray}\label{eq-42}
f({z})=\sqrt{\frac{\mu}{2\pi  D \beta }} \mathrm{exp}\left(- \frac{\mu z^2}{{2 D\beta}} \right), 
\end{eqnarray}
where $\beta = \nu(\mu+\nu)^{-1}$. 

It is found that the coefficient $\beta$ ($=1/(1+\mu/\nu)$) is inversely proportional to correlation time $T_L(=\nu^{-1}) $ of the colored noise.  Therefore, longer correlation time $T_L(=\nu^{-1}) $ of the colored noise leads to weaker diffusion.  Differently, for the cases that $\nu \to \infty$, $\langle \langle\dot{{X}}_e(\tau_1^{\prime}) \dot{{X}}_e(\tau_2^{\prime})\rangle\rangle \to 2D \delta (\tau_1-\tau_2)$, implies it is reduced to a simple white noise, leading to a stronger diffusion.

\begin{figure}[h]
   \centering
   \subfloat[$\mu=1$ and $\nu=0.1$]
   {
   	\label{fig:Fig2:a}
   	\begin{minipage}[c]{0.333333\textwidth}
   		\centering
   		\includegraphics[width=\textwidth]{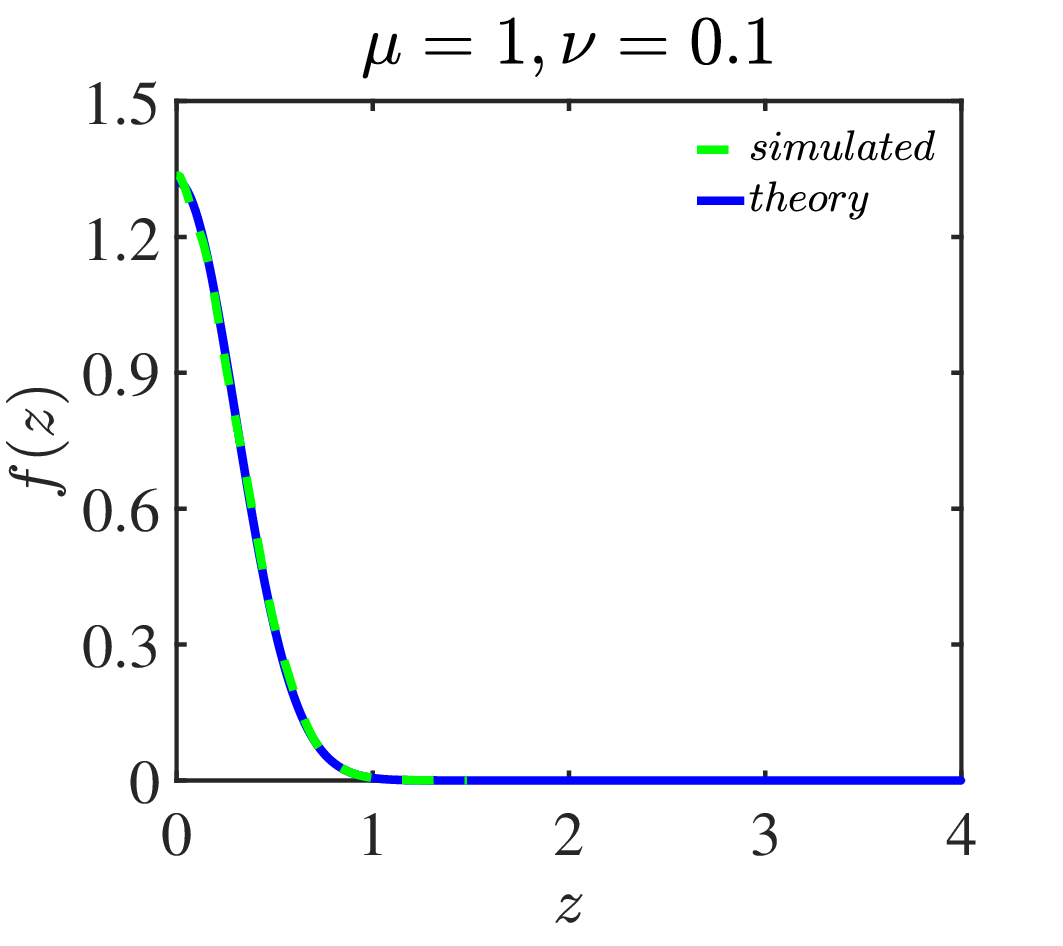}
   	\end{minipage}
   }  
   \subfloat[$\mu=1$ and $\nu=1$]
   {
   	\label{fig:Fig2:b}
   	\begin{minipage}[c]{0.333333\textwidth}
  		 \centering
   		\includegraphics[width=\textwidth]{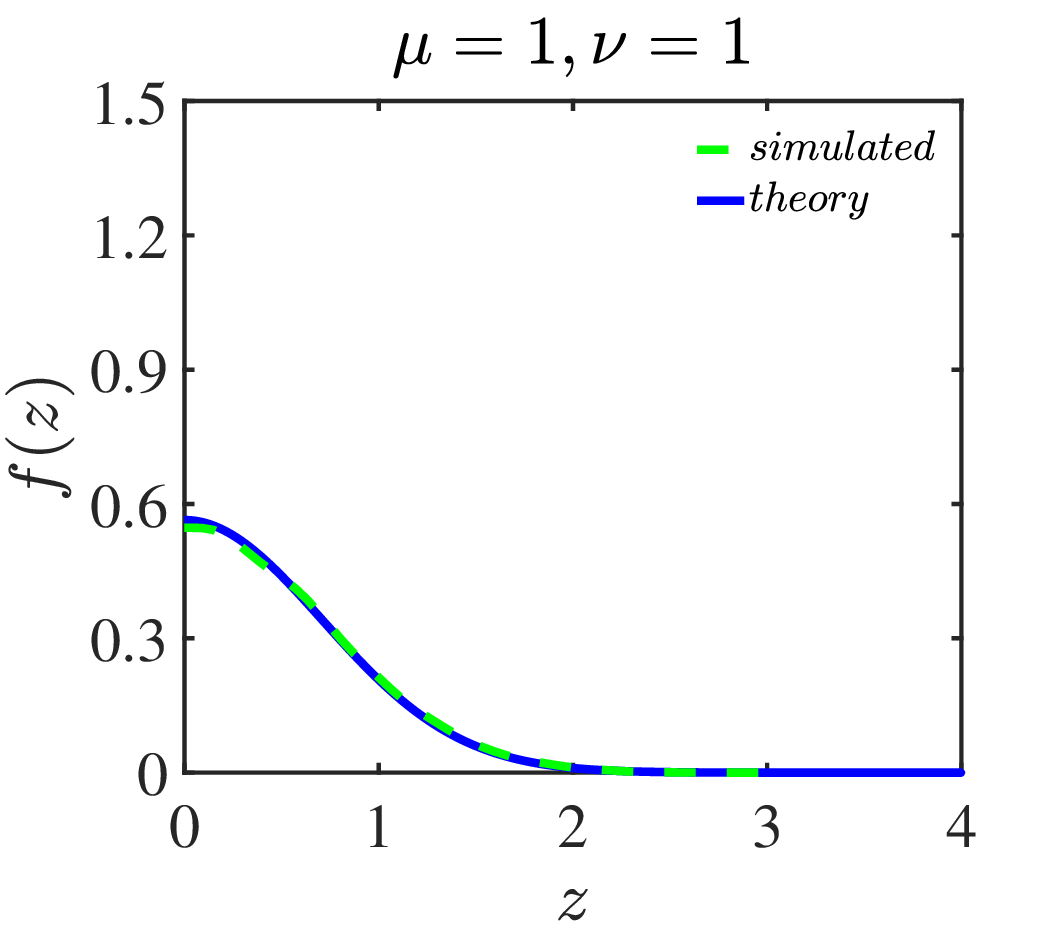}
   	\end{minipage}
   } 
   \subfloat[$\mu=1$ and $\nu=10$]
   {
   	\label{fig:Fig2:c}
   	\begin{minipage}[c]{0.333333\textwidth}
   		\centering
   		\includegraphics[width=\textwidth]{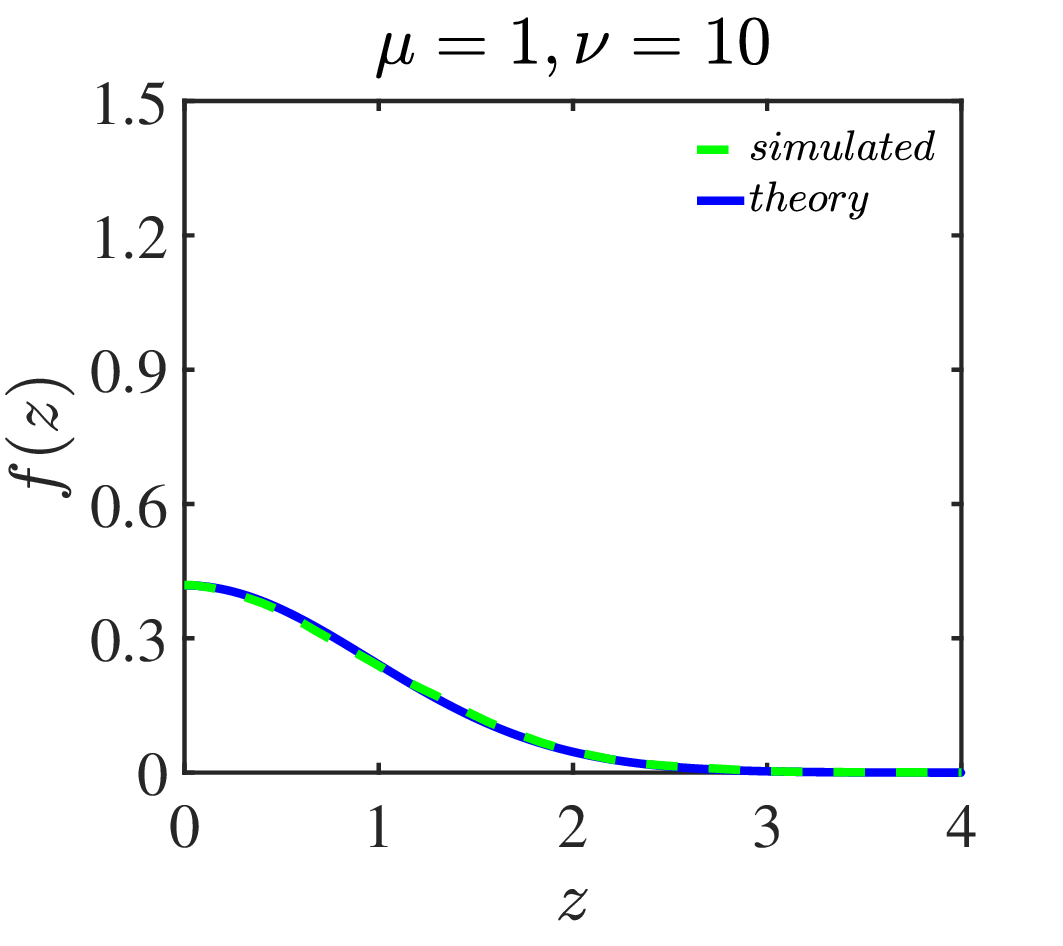}
   	\end{minipage}
   } 
\caption{\emph{Stationary PDF of $z$ given by Eq. \eqref{eq-42}. The distributions of $z$ obtained by numerical solutions of Eq. \eqref{eq-33} are also plotted, of which $\dot{X}_e$ is considered as a colored Gaussian noise given by Eq. \eqref{eq-38}. Since the coefficient $\beta$ ($=1/(1+\mu/\nu)$) is inversely proportional to correlation time $T_L(=\nu^{-1}) $ of the colored noise, longer correlation time $T_L(=\nu^{-1}) $ leads to weaker diffusion.}}
\label{fig:Fig2} 
\end{figure}

Fig. \ref{fig:Fig2} shows the distributions of $z$ given by Eq. \eqref{eq-42} for different $\mu$ and $\nu$. For comparison, the simulated distributions are also plotted Fig. \ref{fig:Fig2}, which are determined by solving Eq. \eqref{eq-33} numerically regarding $\dot{X}_e$ as a colored Gaussian noise (Eq. \eqref{eq-38}) \cite{risken1984fokker}. It shows that Eq. \eqref{eq-42} agrees well  with simulations. 

%
%

\section{\label{sec:level5}Concluding Remarks}

Evolution of a complex system from one state to another can have a large number of possible paths. Moreover, non-uniformity in field variables leads the local dynamics in state transition differs considerably from path to path, which ultimately affects statistical characteristics of the system. This study aimed to provide a kinetic equation for the path-dependent system. Also, because 
long-time correlation is an intrinsic feature of path-dependent systems, this study is expected to be helpful in research regarding stochastic nature of systems affected by memory effect.  

This study started from a fundamental postulation in statistical mechanics. It is said that, a macroscopic state of a system belongs to a statistical ensemble composed of a huge number of different microscopic configurations compatible with their macroscopic constraints, and thus any possible realizations of the macroscopic state, as well as its transition paths in the phase space, exhibits a certain degree of uncertainty. This uncertainty, especially that in state transition paths, is what we focus on and the path-average kinetic equation is derived, rather than randomness due to system noises as in the Langevin equation.

In order to account for the uncertainty in state transition paths, we introduced a local path density function to determine their distribution to constitute a statistical ensemble for state transition paths. Upon this state transition ensemble, we developed a new kinetic equation for path-dependent systems. The kinetic equation is derived by a cumulant expansion with respect to state transition path. The kinetic equation shares the similarity in form to the Kramers-Moyal expansion, but with significant differences. The most obvious one is that the expansion coefficients of the Kramers-Moyal expansion are the functions of jump moments; while in our study, they are functions of cumulants with respect to state transition paths. 
It also shows that, for short-time correlations, the path-averaged kinetic equation is reduced to the Fokker-Planck equation. 

The improvement made in this study enables the present study can be applied to the circumstances where the path-dependence and accompanied long-time correlation play a major part in system evolution. Also, the present study is expected to provide an alternative approach to discuss memory effect which is an essential feature of path-dependent systems.

\bibliographystyle{ieeetr}
\bibliography{bib2019.bib}

\end{document}